\documentclass[12pt,a4paper]{article}

\usepackage{amsmath,amssymb,amsfonts}

\newcommand{\bse}{\begin{subequations}}
\newcommand{\ese}{\end{subequations}}
\newcommand{\be}{\begin{equation}}
\newcommand{\ee}{\end{equation}}
\newcommand{\bea}{\begin{eqnarray}}
\newcommand{\eea}{\end{eqnarray}}
\newcommand{\ba}{\begin{array}}
\newcommand{\ea}{\end{array}}

\newcommand{\p}{\partial}

\def\ZZ{{\mathcal{Z}}}

\def\DD{{\mathcal{D}}}
\def\GG{{\mathcal{G}}}

\input amssym.def
\input amssym.tex

\usepackage[colorlinks=true, linkcolor=blue, bookmarks=true, citecolor=red ]{hyperref}

\usepackage{geometry}
\begin{document}

\title{Gravitational Effective Action: A New Method}
\date{}

\maketitle
\hrule

\linespread{1.5}

\begin{center}
\vspace{5mm}

\author{Davood Allahbakhshi\footnote{allahbakhshi@ipm.ir}\\\vspace{5mm} School of Particles and Accelerators, Institute for Research in Fundamental Sciences (IPM), P.O.Box 19395-5531, Tehran, Iran}

\vspace{5mm}


\vspace{5mm}
\end{center}

\hrule

\abstract{
\noindent
A new method is introduced for doing calculations of quantum field theories in planar geometries which the metric depends on just one coordinate. In contrast to previous method, this method can be used in any planar geometry, not only weakly curved asymptotically flat ones.}

\newpage

\hrule

\tableofcontents

\vspace*{5mm}
\hrule

\section{Introduction}

Investigating quantum phenomena in curved backgrounds has a long, very interesting story started by finding particle production from the viewpoint of non-inertial observers by Fulling \cite{Fulling:1972md}, Davies \cite{Davies:1974th} and Unruh \cite{Unruh:1976db} and also static observer at infinity of a black hole by Hawking \cite{Hawking:1974sw}. Such calculations led to some serious problems such as information paradox \cite{Hawking:1976ra}. Thermality of Hawking radiation which is the direct result of eternity of the black hole under consideration means that there is no information leakage from the black hole. Some physicists believe that the information paradox will be solved if we could include the backreaction of the radiation on the black hole. This backreaction pushes the radiation away from thermality and so the radiation is not generally informationless.
Including the backreaction of the radiation on the black hole has some physical and technical difficulties. The geometry of the space reflects some fraction of the radiation back to the black hole and so all radiated particles do not reach the observer at infinity or boundary of the space, if it exists. this fraction which is named \emph{greybody factor}, generally is related to the frequency of the particle which wants to move through the space, as well as the geometry itself. Also if there is any boundary in the space, the boundary conditions which change the Hilbert space of the system and also some effects like reflection from the boundary to the black hole, should be considered. Although for large black holes, which the Hawking temperature is low and the backreaction of the radiated particles\footnote{In fact the low energy particles.} can be ignored, the Hawking's formula for black hole radiation can be used to a good approximation. But since the backreaction is ignored any non-trivial effect is in fact ignored as well.

Also in early universe cosmology, including the effect of quantum modes on the background metric is very important.

It seems the best way for overcoming all these difficulties is including the effect of quantum matter modes directly into the action. In fact if we can \emph{integrate out} the matter fields, then we will have a \emph{gravitational effective action} with the metric as a classical field which includes all quantum effects of matter fields. Such effective action should produce all interesting phenomena such as particle production in time dependent backgrounds, non-inertial coordinate systems and also black holes with all backreactions. Also the effect of boundary conditions and reflections both from geometry and the boundary, contribution of quantum matter fields in dynamics of the metric during early universe evolutions, etc. will be included in such an effective action.

During 1980's some physicists started moving in this direction. A covariant method proposed for expanding the one-loop effective action in powers of \emph{curvature} for \emph{asymptotically flat metrics} by Barvinsky and Vilkovisky \cite{Barvinsky:1987uw}. Later they calculated the second \cite{Barvinsky:1990up,Barvinsky:1990uq} order terms. Barvinsky, Gusev, Zhytnikov and Vilkovisky also calculated the third order terms in curvature \cite{Barvinsky:1993en}. But by considering higher order terms the calculations become increasingly difficult. On the other hand for highly curved geometries, like black holes, this effective action is not practically usefull because most of the geometries with interesting features can not be written as a weakly curved metric, globally. Also in some areas such as cosmology and gauge/gravity duality, the space is not asymptotically flat.

It is obvious that we need to explore and develop other methods to calculate the gravitational effective action for \emph{not necessarily weakly curved and asymptotically flat} geometries and the aim of present paper is introducing such method.

The paper is organized as follows. In next section we briefly review the concept of functional determinants and transformation between different bases. We also review the functional calculations. In third section we show a simple perturbative calculation of the gravitational effective action for a generic near flat space and its renormalization. In section 4, we introduce \emph{gauge fixing method} in an example. Then we use the method to calculate the first perturbative terms\footnote{In the sense that we will explain later.} of the vacuum expectation value of the energy-momentum tensor and for this we need to renormalize the effective action. Finally we will complete renormalization by going away from the gauge and we will calculate the full effective action which can be used when the gauge is not fixed.

\section{Functional Determinants: A Review}

Functional (path integral) formalism of quantum field theory is a very beautiful and nice formalism in modern physics. Mathematically any calculation in quantum field theory is calculating a functional determinant or its derivatives.
In this section we want to have a brief review on functional calculations in quantum field theory.

Consider a complex scalar field with the action
\be
S = \int{ \big [ \eta^{\mu\nu}\p_\mu \phi(x)^* \p_\nu \phi(x)  - \phi(x)^*\;I(x)\;\phi(x) \big ] \; d^{d+1}x},
\ee
where $I(x)$ can be mass or any external field such as $\zeta \mathcal{R}(x)$\footnote{The Ricci scalar, which produces the non-minimal coupling term of the field to gravity.}. Then the partition function of this theroy is
\be
\ZZ = \int{e^{i S(\phi)} \DD\phi}=\int{e^{i \int{[ \;\p_\mu \phi ^* \p^\mu \phi \; - \; \phi ^* I \phi \;] \;d^{d+1}x}} \DD\phi}.
\ee
At first step by a wick rotation $ t \rightarrow i \tau $, we \emph{Euclideanize} the action
\be
\ZZ = \int{e^{ -\int{ [\; \delta^{\mu\nu}\; \p_\mu \phi ^* \p_\nu \phi \; - \; \phi ^* I \phi \;] \;d^{d+1}x}} \DD\phi}.
\ee
Then after an integration by parts up to a boundary term we have
\be
\ZZ = \int{e^{ \int {[\; \phi ^* \p^2 \phi \; + \; \phi ^* I \phi \; ] \; d^{d+1}x}} \DD\phi},
\ee
where $\p^2$ is $\delta^{\mu\nu}\p_\mu\p_\nu$.

As you know this is the functional determinant of the operator $\p^2 + I$. On the other hand $\p^2$ as an operator is nothing but the inverse of the Green's function of the free theory $\GG^{-1}$. So we have
\be\label{FD1}
\ZZ = Det[\GG^{-1}+I]^{-1}=e^{-tr [ln(\GG^{-1}+I)]}.
\ee
If we can calculate this determinant completely then we are using some non-perturbative method but, if we can not, then we can use \emph{perturbation}.
If $I$ is small in comparison to $\GG^{-1}$, then we can expand the logarithm in the previous equation
\be\label{FD2}
\ZZ = e^{-tr[ln(\GG^{-1})]-tr [ln(1+\GG I)]}=Det[\GG] e^{-tr[\GG I]+\frac{1}{2}tr[\GG I \GG I]+...}.
\ee
So the effective action will be
\be\label{EF1}
\Gamma = \Gamma_0 - tr[\GG I]+\frac{1}{2}tr[\GG I \GG I]+...\;,
\ee
where $\Gamma_0 = tr[ln(\GG)] $, most of the time a constant infinite number.

There is a very important point about this calculation related to the basis in which the trace of the operators are being calculated. In \ref{FD1} the trace is calculated by the spectrum of the operator $\mathcal{O}=\GG^{-1}+I$ but, in \ref{FD2} the trace is calculated by the spectrum of $\GG^{-1}$. Let us see this point in detail.
Suppose that the spectrums of $\mathcal{O}=\GG^{-1}+I$ and $\GG^{-1}$ are $(-\Lambda_n^2,\psi_n)$ and $(-\lambda_n^2,\phi_n)$, enjoy orthonormality conditions
\bea
\int{\psi_m(x)^*\psi_n(x)\;d^{d+1}x}=\delta_{mn}\cr
\int{\phi_m(x)^*\phi_n(x)\;d^{d+1}x}=\delta_{mn}.
\eea
These spectrums are the set of eigenfunctions which make the operators $\mathcal{O}$ and $\GG^{-1}$ diagonal respectively. then we have
\be
\ZZ =\int{e^{-\Lambda_n^2 \; |c_n|^2}\prod_k dc_k dc^*_k}= \prod_n \frac{\pi}{\Lambda_n^2}.
\ee
But in practice we do not have the spectrum of $\mathcal{O}$ and we want to use the spectrum of $\GG^{-1}$ to calculate the path integral. For this we can write

\bea\label{trans1}
\ZZ &&= \int{e^{c^*_m c_n\int{\psi^*_m \mathcal{O} \psi_n\; d^{d+1}x}}\prod_k dc_k dc^*_k}\cr
&&=\int{e^{\hat{c}^*_m \hat{c}_n\int{\phi^*_m (\GG^{-1} + I) \phi_n\; d^{d+1}x}}\prod_k dc_k dc^*_k}.
\eea
In the path integrals above $\hat{c}_n$ is
\be
\hat{c}_n = c_m \; J_{mn},
\ee
where $J_{mn}$ is the Jacobian of the transformation from the spectrum $\phi$ to the spectrum $\psi$
\be\label{jacobian}
\psi_m = J_{mn}\;\phi_n \rightarrow J_{mn}=\int{\psi_m \; \phi^*_n \; d^{d+1}x}.
\ee
It is so easy to see that this transformation is unitary since both the basis $\psi_n$ and $\phi_n$ are \emph{orthonormal} and \emph{complete}.
\bea
&&\phi_m = J^{-1}_{mn}\;\psi_n \rightarrow J^{-1}_{mn}=\int{\phi_m \; \psi^*_n \; d^{d+1}x}=J_{nm}^*\cr
&&\Rightarrow J^{-1}=J^\dagger.
\eea
Note that in second line in \ref{trans1} we have kept the measure of the path integral as in the first line. Also we need to change the measure and the important point is here. The measure can be transformed as
\be
\prod_n dc_n dc^*_n = Det(J^\dagger)Det(J) \prod_n d\hat{c}_n d\hat{c}^*_n=Det(J^\dagger J) \prod_n d\hat{c}_n d\hat{c}^*_n.
\ee
Since the transformation $J$ is unitary, the measure is invariant. If the measure was not invariant then we had another functional determinant ($Det(J^\dagger J)$) which generally depends on the spectrum $\psi_n$ that we do not khow. Now thanks to this invariancy we can calculate the path integrals perturbatively.

Let us see the perturbative calculation of the partition function directly. At first we expand the interaction term

\be\label{pathintegral1}
\ZZ = \int{e^{\int \phi^* \GG^{-1} \phi}\big[ 1-\int \phi^* I \phi \; d^{d+1}x + ... \big] \DD\phi }.
\ee
From now on we drop the integrals over coordinates, for simplicity. We can expand any field $\phi(x)$ in terms of the eigenfunctions of $\GG^{-1}$
\be
\phi(x)=\sum_n{c_n \phi_n(x)}.
\ee
The first term in partition fuction \ref{pathintegral1} will be
\be
e^{\Gamma_0} = \int{e^{-\lambda_i^2 |c_i|^2}\prod_k dc_k dc^*_k}=\prod_i \frac{\pi}{\lambda_i ^2}.
\ee
The second term is
\be\label{pathintegral2}
\sum_{m,n}\int{e^{-\lambda_i^2 |c_i|^2}  c^*_m c_n I_{mn} \prod_k dc_k dc^*_k},
\ee
where
\be 
I_{mn}=\int{\phi^*_m(x) I(x) \phi_n(x) d^{d+1}x}.
\ee
The path integral \ref{pathintegral2} is the product of the integrals for every mode. If $m\neq n$ it is
\be
\sum_{m,n}\;I_{mn}\;\bigg(\int{e^{-\lambda_1^2 |c_1|^2} dc_1 dc^*_1}\bigg)...\bigg(\int{e^{-\lambda_m^2 |c_m|^2} c^*_m dc_m dc^*_m}\bigg)...\bigg(\int{e^{-\lambda_n^2 |c_n|^2} c_n dc_n dc^*_n}\bigg)...
\ee
The integrals of $dc_m$ and $dc_n$ are zero since their integrands are odd functions. The result is non-zero just when $m=n$ and it is
\bea\label{pathintegral3}
&&\sum_{m,n} I_{mn} \delta_{m,n} \int{e^{-\lambda_i^2 |c_i|^2}  |c_m|^2 \prod_k dc_k dc^*_k}=e^{\Gamma_0} \times \sum_{n} \frac{I_{nn}}{\lambda_n^2} =\cr &&e^{\Gamma_0} \times \int{I(x)\sum_{n} \frac{\phi^*_n(x)\phi_n(x)}{\lambda_n^2}d^{d+1}x}.
\eea
The last sum is nothing but the eigenfunction expansion of the Green's function. So the result is
\be\label{tdpole1}
-e^{\Gamma_0} \times \int{I(x)\GG (x,x) \; d^{d+1}x}.
\ee

The same calculation for the next term in \ref{pathintegral1} leads to
\be
e^{\Gamma_0} \times \frac{1}{2}\sum_{m,n,p,q}\;\frac{I_{mn}I_{pq}}{\lambda_m \lambda_n \lambda_p \lambda_q }(\delta_{mn}\delta_{pq}+\delta_{mq}\delta_{np}),
\ee
which is
\be
e^{\Gamma_0} \times \bigg [\frac{1}{2}\int{\GG (x,y)I(y)\GG (y,x)I(x)\; d^{d+1}x}+\frac{1}{2}\big(\int{\GG (x,x)I(x)\; d^{d+1}x}\big)^2 \bigg].
\ee
These terms are in fact all possible contractions of the fields in the canonical formalism. The second term in the expression above is the square of \ref{tdpole1}. It is well known that higher orders of this term (as well as other terms like the first term in the result above and also higher order terms) appear in next terms in such a way that the final result will be
\be
\ZZ = e^{\Gamma_0 + connected\;\;terms}.
\ee
So
\be\label{EF2}
\Gamma = \Gamma_0 - \int{\GG (x,x)I(x)\; d^{d+1}x} + \frac{1}{2} \int{\GG (x,y)I(y)\GG (y,x)I(x)\; d^{d+1}x} + ...,
\ee
which is nothing but \ref{EF1}. The core concept in all these calculations is the existence of an operator $\GG^{-1}$ that we know its spectrum exactly and so we can calculate the functional trace by the use of this spectrum.

\section{An Introductory Example}

In this section we show a simple calculation of the first term of the gravitational effective action and the usual way that we renormalize the result.

\subsection{The First Term}

Consider a scalar field which lives in a weakly curved spacetime. Since the space is weakly curved its metric can be written as a small deviation from the flat space metric
\be\label{metricvar}
g_{\mu\nu} = \eta_{\mu\nu} +h_{\mu\nu} .
\ee
The action of the scalar field to first order in $h_{\mu\nu}$ is
\be
S = \int{\big[\;\p_\mu\phi^* \p^\mu\phi \;+\; X^{\mu\nu}\p_\mu \phi^* \p_\nu \phi \;\big]\;d^{d+1}x},
\ee
where $X^{\mu\nu}$ is
\be
X^{\mu\nu}=h^{\mu\nu}-\frac{1}{2} \; h_\rho^\rho \; \eta^{\mu\nu}.
\ee
The first term of the effective action can be calculated by the same method used in the previous part and the result is
\be
\int{X^{\mu\nu}\frac{\p}{\p x^\mu}\frac{\p}{\p y^\nu} \GG (x,y)\big|_{y=x}\;d^{d+1}x}.
\ee
By integrating by parts the above expression can be written as\footnote{Note that we have assumed the boundary terms which arise after integrating by parts are ignorable.}
\be
\int{\GG (x,x)\;\p_\mu\p_\nu X^{\mu\nu}\;d^{d+1}x}.
\ee
As you know $\GG (x,x)$ is an infinite constant number. Also $\p_\mu\p_\nu X^{\mu\nu}$, up to a total derivative, is the Ricci scalar to first order in $h^{\mu\nu}$
\be
\p_\mu\p_\nu X^{\mu\nu} \approx \sqrt{-g}\;\;\mathcal{R}.
\ee
So
\be
\Gamma(g) \approx \Gamma_0 + \infty \times \int{\mathcal{R}\;\sqrt{-g}\;d^{d+1}x}+...
\ee

\subsection{Renormalization}

For renormalizing the previous result we suppose that there is also a gravitational action (Einstein-Hilbert action e.g.) which includes $\mathcal{R}$
\bea
\Gamma_{total}(g) &&= \frac{1}{4\pi G_{N-bare}}\int{\mathcal{R}\;\sqrt{-g}\;d^{d+1}x}+\Gamma (g) \cr\cr
&&= \Gamma_0 + \frac{1}{4\pi G_{N-physical}}\int{\mathcal{R}\;\sqrt{-g}\;d^{d+1}x} + ...
\eea
In the second line we have absorbed infinity into $G_{N-bare}$ and introduced $G_{N-physical}$.
Second and third orders of this calculation are also done and the result is a series in curvature named \emph{Seeley-DeWitt} expansion in addition to some non-local terms. These higher order terms in the gravitational effective action also have infinite coefficients which can be absorbed in similar terms which are supposed to exist in gravitational action\footnote{Be careful about the terms \emph{gravitational action} and \emph{gravitational effective action}. Gravitational action is some classical action such as Einstein-Hilbert action and gravitational effectve action is $\Gamma (g)$.}.

\section{Gauge Fixing Method}

In previous section we calculated the first term of the effective action for a generic weakly curved asymptotically flat geometry. What if the space is not weakly curved and/or asymptotically flat? many interesting phenomena may occur in highly curved spaces which their metric can not be written in the form of \ref{metricvar}, such as black holes. On the other hand asymptotically de Sitter and Anti-de Sitter geometries are very important in cosmology and gauge/gravity duality. In this section we introduce our method for calculating gravitational effective action for planar geometries, in which the metric can be written as a function of just one coordinate. We name this method, the \emph{gauge fixing method}. There are many important and interesting planar geometries including Rindler space, flat Friedmann-Robertson-Walker metric, Anti-de Sitter space in Poincare coordinate, black branes, etc.

Consider a curved space equipped with a metric $g_{\mu\nu}$ and suppose that there is a series of frames in which the metric is a function of just one coordinate $z$. We want to calculate the effective action in this series of frames. This is the first part of our gauge fixing. In such frames the action of a complex scalar field is
\be
S = \int{\sqrt{-g(z)}\big[g^{zz}\p_z\phi^*\p_z\phi + g^{z\mu}(\p_z\phi^*\p_\mu\phi + \p_\mu\phi^*\p_z\phi) + g^{\mu\nu}\p_\mu\phi^*\p_\nu\phi\big]dz\;d^dx}.
\ee
The next step is changing the coordinate $z$ to $\xi$ in a way that $\sqrt{-\tilde{g}(\xi)}\;\tilde{g}^{\xi\xi} = 1$. This is the second part of our gauge fixing. This coordinate transformation can be found from the relation
\be
d\xi = \frac{dz}{\sqrt{-g(z)}\;g^{zz}} \Rightarrow \xi = \int{\frac{dz}{\sqrt{-g(z)}\;g^{zz}}}
\ee
and the action will be
\be
S = \int{\big[\p_\xi\phi^*\p_\xi\phi + A^\mu(\p_\xi\phi^*\p_\mu\phi + \p_\mu\phi^*\p_\xi\phi) + B^{\mu\nu}\p_\mu\phi^*\p_\nu\phi\big]d\xi\;d^dx},
\ee
where
\bea
\sqrt{-\tilde{g}}\;\tilde{g}^{\xi\xi} &&= 1\cr\cr
A^\mu = \sqrt{-\tilde{g}}\;\tilde{g}^{\xi\mu} &&= \sqrt{-g}\;g^{z\mu}\cr\cr
B^{\mu\nu} = \sqrt{-\tilde{g}}\;\tilde{g}^{\mu\nu} &&= det(g)\; g^{zz}\;g^{\mu\nu}.
\eea
Since both $A^\mu$ and $B^{\mu\nu}$ are functions of $\xi$, eigenfunctions of this action are of the form
\be
\phi(\xi , x)=\int{\phi(\xi,k)\;e^{ik.x}d^dk}.
\ee
So the action is
\be
S = \int{\big[\p_\xi\phi^*\p_\xi\phi + A \; (\phi \; \p_\xi\phi^* - \phi^*\p_\xi\phi) + B \; \phi^*\phi\big]d\xi\;d^dk},
\ee
where
\be
A=i A^\mu k_\mu  \;,\; B = B^{\mu\nu}k_\mu k_\nu .
\ee
It is obvious that we can turn everything back to the position space by using the relations
\bea
&&\int{i k_\mu \;d^dk}=\int{\delta(x)\p_\mu \delta(x) \;d^dx}\cr\cr
&&\int{k_\mu k_\nu \;d^dk}=\int{\p_\mu\delta(x)\p_\nu \delta(x) \;d^dx}.
\eea
Note that $B$ also can include the \emph{mass} and \emph{non-minimal coupling} of the field to gravity
\be
B = B^{\mu\nu}k_\mu k_\nu + \frac{1}{g^{\xi\xi}}(m^2 +\zeta \mathcal{R}).
\ee
After integrating the first term in the action by parts, we have
\be
S = \int{\big[-\phi^*\p^2_\xi\phi + A \; (\phi \; \p_\xi\phi^* - \phi^*\p_\xi\phi) + B \; \phi^*\phi\big]d\xi\;d^dk}.
\ee
By a wick rotation $\omega \rightarrow i\omega $, we have the Euclidean action
\be\label{FA1}
S= i S_E = i\; \int{\big[-\phi^*\p^2_\xi\phi + \hat{A} \; (\phi \; \p_\xi\phi^* - \phi^*\p_\xi\phi) + \hat{B} \; \phi^*\phi\big]d\xi\;d^d\hat{k}},
\ee
where $\hat{A},\hat{B}$ and $d^d\hat{k}$ are the Euclidean functions and measure. For simplicity we drop this hat sign. We want to calculate the path integral
\bea
\ZZ &&= \int{e^{iS}\DD\phi}=\int{e^{-S_E}\DD\phi}\cr
&&=\int{e^{\int{\big[\phi^*\p^2_\xi\phi \; - A \; (\phi \; \p_\xi\phi^* - \phi^*\p_\xi\phi) - B \; \phi^*\phi\big]d\xi\;d^dk}}\DD\phi}.
\eea
Now we are ready to calculate this functional determinant. We consider $\p_\xi ^2$ as the $\GG^{-1}$ and other terms as interaction terms\footnote{Note that we consider the action as a \emph{one-dimensional} theory with $d$ \emph{internal degrees of freedom} $k_\mu$.} and use the spectrum of $\GG^{-1}$ to calculate the path integral. The spectrum of $\p_\xi ^2$, without any special boundary condition, is obviously $(-q^2,e^{iq\xi}/\sqrt{2\pi})$ and as mentioned in previous section the transformation from complete basis to this basis is unitary and so the measure of the path integral is invariant. The terms $A$ and $B$ will be our interaction terms and the calculation will be perturbative in these terms. But there are two important questions: \emph{In which sense it is a perturbation?} In the absence of the mass and non-minimal coupling to gravity, the terms $A$ and $B$ include the momentum $k_\mu$ that we name \emph{transverse momentum}. So this perturbation is a perturbation in transverse momentums or in other words it is a \emph{hydrodynamical expansion in transverse planes}. The second question is this \emph{Is this perturbation useful?} In some cases we are interested in the \emph{contribution of every momentum} to some quantity such as energy-momentum tensor instead of the full quantity itself. Although there is an integral over the transverse momentums, it can be seen directly in calculations that since the metric is just function of $\xi$ and the field theory is non-interactive, there is not any mixing between transverse momentums so the final result will have an overal integral over these momentums. It means we can calculte any quantity by dropping the integral over transverse momentums and include the integral just in final result. So by this method we can calculate the contribution of low transverse momentums to any quantity perturbatively.

The most important objects appear in our calculations are the Green's function and its derivatives. As mentioned in previous section, the Green's function appears in the form of an eigenfunction expansion. In the present case the spectrum is $(- q^2,e^{iqx}/\sqrt{2\pi})$, so the \emph{one dimensional} Green's function is
\be\label{gf1}
\GG (x,y)=-C[\phi^*(x),\phi(y)] = - \frac{1}{2\pi}\int{\frac{e^{-iq(x-y)}}{q^2}dq},
\ee
where $C[\bullet,\bullet]$ refers to the \emph{contraction} of fields. This Green's function is divergent! so we can not use this function directly in our calculations. There are two possible solutions to this problem. Introducing a \emph{fake mass} and using the \emph{Fourier transformation}.\\\\
- \emph{fake mass}\\
by adding the term $m^2\;\phi^*\phi$ to \ref{FA1} we can regularize the Green's function. We name this \emph{non-physical} mass, \emph{the fake mass}.
\bea
\GG (x,y) = - \frac{1}{2\pi}\int{\frac{e^{-iq(x-y)}}{q^2+m^2}dq}=-\frac{e^{-m|x-y|}}{2m}.
\eea
Obviously this function is divergent at the limit $m\rightarrow 0$, but after it is contracted with the interaction terms $A$ and $B$, the result can be finite. In the same way we have
\bea
C[\p_x\phi^*(x),\phi(y)]&&=-C[\phi^*(x),\p_y\phi(y)]=-\frac{1}{2}\;sign(x-y)\;e^{-m|x-y|}\cr
C[\p_x\phi^*(x),\p_y\phi(y)]&&=\delta(x-y)-\frac{m}{2}e^{-m|x-y|}.
\eea
These functions have finite $m\rightarrow 0$ limits
\bea
C[\p_x\phi^*(x),\phi(y)]&&=-\frac{1}{2}\;sign(x-y)\cr
C[\p_x\phi^*(x),\p_y\phi(y)]&&=\delta(x-y).
\eea
Also you can check these results from the original definition \ref{gf1}.\\\\
- \emph{Fourier transformation}\\
Instead of calculating the Green's function in position space we can calculate the Fourier transform of the interaction terms
\bea
I(q,p)=\int{\big[B(x)-i(p+q)A(x)\big]e^{i(p-q)}dx},
\eea
and then a typical calculation will be
\be
\int{\big[... \frac{1}{q^2} I(q,p) \frac{1}{p^2}I(p,k) ...\big] \; dpdqdk...}\;.
\ee
Again depending on the interaction terms, the result can be finite.

The fake mass method is practically simple and we can work explicitly with the metric as a classical field in space-time, although the Fourier transformation method is also good because summarizes all terms in just one momentum space object $I(q,p)$, and the form of any quantity is trivially known.

\subsection{A Simple Calculation}
Let us calculate the effective action to second order in transverse momentums ($O(k^2)$). The first terms we need to calculate are
\bea\label{myeff1}
&&-\int{e^{\phi^* \p_\xi^2 \phi}\big[ A \; (\phi \; \p_\xi\phi^* - \phi^*\p_\xi\phi) + B \; \phi^*\phi \big] \DD \phi}=\cr\cr
&&-\int{A(x)\big[ C[\phi(x),\p_x \phi(x)^*] - C[\phi(x)^*,\p_x \phi(x)] \big]\;dx}-\int{B(x)\GG (x,x)\;dx}=\cr\cr
&&2i \bigg(\int_{-\infty}^{+\infty}{\frac{dq}{q}}\bigg) \bigg(\int{A(x)\;dx}\bigg) - \bigg(\int_{-\infty}^{+\infty}{\frac{dq}{2\pi q^2}}\bigg) \bigg(\int{B(x)\;dx}\bigg).
\eea
In the first line we have dropped the integral over $x$, for simplicity. The integral $\int dq/q$ is zero since the integrand is an odd function of $q$. The second integral $\int dq/q^2$ is divergent and needs to be renormalized. In next section we will renormalize the effective action in an appropriate way.

The next term we need to calculate is
\bea\label{myeff2}
&&\frac{1}{2}\int{e^{\phi^* \p_\xi^2 \phi}\big[ - A \; (\phi \; \p_\xi\phi^* - \phi^*\p_\xi\phi)\big] \big[ - A \; (\phi \; \p_\xi\phi^* - \phi^*\p_\xi\phi)\big] \DD \phi} = \cr\cr
&&  \int{A(x)A(y)\big(-\frac{1}{2}Sign(x-y)\big) \big(-\frac{1}{2}Sign(y-x)\big) \;dxdy}+\cr\cr
&&  \int{A(x)A(y)\big(\GG (x,y) \delta(x-y)\big) \;dxdy} =\cr\cr
&& -\frac{1}{4}\bigg(\int{A(x)\;dx}\bigg)^2-\bigg( \int_{-\infty}^{+\infty}{\frac{dq}{2\pi q^2}} \bigg)\int{A(x)^2\;dx}.
\eea
Again the same divergency has appeared. Note that in all these calculations, as mentioned previously, there is an overall integral over transverse momentums that we have dropped for simplicity. So the effective action to second order in transverse momentums is
\bea
\Gamma(g)&&=\Gamma_0 - \bigg(\int_{-\infty}^{+\infty}{\frac{dq}{2\pi q^2}}\bigg) \int{\big[ B(x) + A(x)^2 \big]\;dx}\cr\cr && -\frac{1}{4}\bigg(\int{A(x)\;dx}\bigg)^2.
\eea
It is interesting that, to this order, all terms are local. In the next section we calculate the vacuum expectation value of energy-momentum tensor and we need to renormalize the effective action.

\subsection{Energy-Momentum Tensor}
For calculating the energy-momentum tensor we need to relax the gauge fixing condition $\sqrt{-g}\;g^{\xi\xi} = 1$, vary the metric and calculate the coefficient of the variation which is the vacuum expectation value of the energy-momentum tensor. Let us name $X=\sqrt{-g}\;g^{\xi\xi}$ for simplicity. Under variation of the metric the interaction terms also vary. we name them $\delta A, \delta B$ and $\delta X$.
\be
\delta S_E = \int{\big[ \delta X\;\p_\xi \phi^* \p_\xi\phi + \delta A(\phi \; \p_\xi \phi^* - \phi^* \p_\xi \phi) + \delta B \phi^* \phi \big]d\xi d^dk}.
\ee
It is obvious that for calculating the effective action with these variations, it is enough to calculate the path integral with the action
\be
S_E = \int{\big[ (1+\delta X)\;\p_\xi \phi^* \p_\xi\phi + A(\phi \; \p_\xi \phi^* - \phi^* \p_\xi \phi) + B \phi^* \phi \big]d\xi d^dk}.
\ee
We have calculated the effective action to second order in transverse momentums previously without $\delta X$. The extra terms which should be calcuated to second order in transverse momentums are\footnote{We have dropped $\p_M \phi^* \p_N \phi$ for simplicity.}
\bea
-&&\int{e^{\phi^* \p^2_\xi \phi}\delta X\;\DD \phi}\cr
+&&\int{e^{\phi^* \p^2_\xi \phi}\delta X\;A\;\DD \phi}\cr
+&&\int{e^{\phi^* \p^2_\xi \phi}\delta X\;B\;\DD \phi}\cr
-\frac{1}{2}&&\int{e^{\phi^* \p^2_\xi \phi}\delta X\;A\;A\;\DD \phi}.
\eea
The first term is
\be
-\int{\delta X \; C[\p_\xi \phi^*, \p_\xi\phi]dx d^dk}=-\delta(0)\int{\delta X \;dx d^dk}.
\ee
It has a divergency ($\delta(0)$) and should be renormalized. The second term is
\bea
&&-\frac{1}{2}\int{\delta X(x)\;A(y)\;Sign(x-y)\;\delta(x-y)\;dxdyd^dk}\cr
&&+\frac{1}{2}\int{\delta X(x)\;A(y)\;Sign(x-y)\;\delta(x-y)\;dxdyd^dk} = 0.
\eea
The third term is
\bea
&&\int{\delta X(x) B(y)\big( -\frac{1}{2}Sign(x-y)\big)\big( -\frac{1}{2}Sign(x-y)\big)\; dxdyd^dk}=\cr\cr
&&\frac{1}{4}\int{\delta X(x) B(y)\; dxdyd^dk}.
\eea
And you can verify that the fourth term is
\be
\frac{1}{2}\int{\delta X(x)\;A(y)\;A(z)\;\Delta(x,y,z)\;dxdydzd^dk},
\ee
where
\be
\Delta(x,y,z)=\delta(x-y)+\frac{1}{2}\delta(y-z)-2\GG(x,x)\delta(x-y)\delta(x-z)
\ee
and the result is
\bea
&&\frac{1}{2} \int{\delta X(x)\;A(x)\;A(y)\;dxdyd^dk}\cr\cr
+&&\frac{1}{4}\int{\delta X(x)\;A(y)^2\;dxdyd^dk}\cr\cr
+&&\bigg(\int_{-\infty}^{+\infty}{\frac{dq}{q^2}}\bigg) \int{\delta X(x)\;A(x)^2\;dxd^dk}.
\eea
And you can check that in total we have
\bea
\Gamma(g) = &&\Gamma_0 - \delta(0)\int{\delta X \;dx d^dk}+\frac{1}{2}\int{\delta X(x)\;A(x)\;A(y)\;dxdyd^dk}\cr\cr
&&+\frac{1}{4}\int{\delta X(x)\;\big[A(y)^2+B(y)\big]\;dxdyd^dk}\cr\cr
&&- \bigg(\int_{-\infty}^{+\infty}{\frac{dq}{q^2}}\bigg) \int{\big[-\delta X(x)\;A(x)^2 + B(x) + A(x)^2 \big]\;dxd^dk}\cr\cr
&& -\frac{1}{4}\int{\bigg(\int{A(x)\;dx}\bigg)^2\;d^dk}.
\eea
In the result above we have two divergencies; $\delta (0)$ in the first divergent term and $\GG (x,x)$ in the second divergent part. Let us start by renormalizing the second divergency which is simpler to be renormalized.\\
There are two important points about this term. The first one is that this divergency exists in the effective action itself, not just in the variation terms. The second point is that in this term we have the combination of $A$ term and $\delta X$ term. These two points mean that we need to add a counter term to the gravitational action to renormalize the effective action as well as its variation. For doing this we note that $X=1$, and also we need to vary the $A$ and $B$ terms as well, so we have
\be
\delta X\;A^2 - 2\;A\;\delta A - \delta B =-\big[\delta\big( (2-X) A^2 \big) +\delta B \big]_{X=1}.
\ee
So the term we should add to gravitational action is
\be
\hat{C}_2\int{\big[(2-X) A^2 + B\big]dxd^dk},
\ee
where $\hat{C}_2$ is a \emph{bare coupling}. This term renormalizes both the effective action and its variation to \emph{first order}; and after fixing the gauge $X=1$, it produces the correct term $A^2$.

Now it is time to renormalize the first divergency. The important point about this term is that this divergency appears just in the variation but not in the effective action itself. It means that we need a counter term that its value is zero but its variation is not! Adding a term $C_1X$ to the action is not good. because although it renormalizes the variation but its value at $X=1$ is not zero. In fact this divergent term is just the first one in a series of divergent terms that their sum is a function with needed features! To see this let us first introduce these terms. They are the terms that are made just by $\delta X$ terms. A typical one is
\be
\int{e^{\phi^* \p_\xi^2 \phi} \big(  \frac{(-1)^n}{n!} \delta X(x_1)...\delta X(x_n)[\p_{x_1}\phi^*\p_{x_1}\phi]...[\p_{x_n}\phi^*\p_{x_n}\phi]  \big)\; \DD \phi }.
\ee
There are $(n-1)!$ ways of contracting the fields, all with the same value. Contracting the fields produce $n$ number of Dirac delta functions. $n-1$ Dirac deltas can be used for identifying the $x_i$s and the final one will be $\delta(0)$. So the result is
\be
\frac{\delta (0)}{n}\int{\big(-\delta X(x)\big)^n\;dxd^dk}.
\ee
The sum of these terms is
\be
-\delta (0)\int{ln(1+\delta X)\;dxd^dk}.
\ee
Which is
\be
-\delta (0)\int{ln(X+\delta X)\big |_{X=1}\;dxd^dk}.
\ee
So by adding the term
\be
\hat{C}_1\int{ln(X)\;dxd^dk}
\ee
to the gravitational action, we can renormalize all these divergencies to \emph{all orders} and interestingly its value, without variation, at $X=1$, is zero, as we needed. In this term $\hat{C}_1$ is another \emph{bare coupling}.

The \emph{renormalized total effective action} is
\bea
\Gamma_{(r-tot)}(g) = &&\Gamma_0 +\int{\frac{1}{\kappa^2}\big(\mathcal{R}-2\Lambda\big)d^{d+1}x}\cr\cr
&&+ C_1\int{ln(X(x))\;dx d^dk}\cr\cr
&& + C_2 \int{\big[(2-X(x)) A(x)^2 + B(x)\big]\;dxd^dk}\cr\cr
&& -\frac{1}{4}\int{\bigg(\int{A(x)\;dx}\bigg)^2\;d^dk}
\eea
and its variation is
\bea
\delta \Gamma_{(r-tot)}(g)=&&\int{\frac{1}{\kappa^2}\delta g^{\mu\nu}\big(G_{\mu\nu}+\Lambda\; g_{\mu\nu}\big)\;\sqrt{-g}\;d^{d+1}x}\cr\cr
&& + C_1\int{ln(1+\delta X(x))\;dx d^dk}\cr\cr
&& + C_2 \int{\big[\delta \big((2-X(x))A(x)^2\big) + \delta B(x)\big]\;dxd^dk}\cr\cr
&& + \frac{1}{2}\int{\delta X(x)\;A(x)\;A(y)\;dxdyd^dk}\cr\cr
&& + \frac{1}{4}\int{\delta X(x)\;\big[A(y)^2+B(y)\big]\;dxdyd^dk}\cr\cr
&& -\frac{1}{2}\int{A(x)\delta A(y)\;dxdyd^dk},
\eea
where $G_{\mu\nu}$ is the \emph{Einstein tensor} and $C_1$ and $C_2$ are \emph{renormalized couplings}. Note that $\delta \Gamma$ is not the variation of $\Gamma$, since in $\Gamma$ the gauge is fixed ($X=1$). In next subsection we will remove this inconsistency by relaxing the gauge and going away from it.

You can verify that the energy-momentum tensor is
\be
T_{AB}=H_{AB}-\frac{1}{2}\big( g^{MN}H_{MN} \big)g_{AB},
\ee
where $H_{AB}$ is
\bea
H_{\xi\xi}(x)&&=C_1-C_2A(x)^2+\frac{1}{2}A(x)\int{A(y)dy}+\frac{1}{4}\int{\big[A(y)^2+B(y)\big]dy}\cr\cr
H_{\xi\mu}(x)&&=i\;\big[C_2A(x) - \frac{1}{4}\int{A(y)dy}\big]k_\mu\cr\cr
H_{\mu\nu}(x)&&=C_2\;k_\mu k_\nu .
\eea

\subsection{Going Away From The Gauge}
In this subsection we want to go away from the gauge $\sqrt{-g}\;g^{\xi\xi}=1$ by a not necessarily small variation $\delta(\sqrt{-g}\;g^{\xi\xi})=\delta X$. We calculte the terms to all orders in $\delta X$ but, the second order in transverse momentums. Finally we read the result as a function of $X$. Let us start from $A$-term. It is
\bea
-\sum_{n=1}^{\infty}{\frac{(-1)^n}{n!}\int{e^{\phi^*\p_\xi^2\phi}\delta X^n A\;[ \p_\xi\phi^*\p_\xi\phi ]...[ \phi \; \p_\xi\phi^* - \phi^*\p_\xi\phi ] \;\DD \phi}}.
\eea
But this term is zero since contraction of $A$ fields with $\delta X$ fields always produce $Sign(0)$ which is zero.

Now let us calculate the $B$-term which is simpler
\bea
-\sum_{n=1}^{\infty}{\frac{(-1)^n}{n!}\int{e^{\phi^*\p_\xi^2\phi}\delta X^n B\;[ \p_\xi\phi^*\p_\xi\phi ]...[ \phi^*\phi ] \;\DD \phi}}.
\eea
All different combinations of the fields produce the same result and the number of combinations is
\bea
2 \times \left( \begin{array}{c}
n \\ 2
\end{array}\right)(n-2)!=n!
\eea
And you can simply check that the value is
\be
-\frac{(-1)^n}{4}\int{\delta X(x)^nB(y)\;dxdy}.
\ee
And so the sum will be
\bea
-\frac{1}{4}\int{\big[ \frac{1}{1+\delta X(x)}-1 \big]B(y)\;dxdy}.
\eea
Note that since we have included all orders of the variation, the function of $X$ should appear in the form of $F(X+\delta X)$. So the result is
\be
-\frac{1}{4}\int{\big[ \frac{1}{X+\delta X(x)}-1 \big]_{X=1}B(y)\;dxdy}.
\ee
It means, without variation, we have
\be
-\frac{1}{4}\int{\big[ \frac{1}{X(x)}-1 \big]B(y)\;dxdy}.
\ee
The final term we need to calculate is $A^2$-term
\bea
\sum_{n=1}^{\infty}{\frac{1}{2}\frac{(-1)^n}{n!}\int{e^{\phi^*\p_\xi^2\phi}\delta X^n A^2\;[ \p_\xi\phi^*\p_\xi\phi ]...[ \phi \; \p_\xi\phi^* - \phi^*\p_\xi\phi ]^2 \;\DD \phi}}.
\eea
There are two patterns of contraction for every term of the sum above
\bea
&&pattern\;1:\;\delta X ... \delta X\;A\;A\cr\cr
&&pattern\;2:\;\underbrace{\delta X ... \delta X}_{i}\;A\;\underbrace{\delta X ... \delta X}_{j=n-i}\;A.
\eea
For pattern 1 the number of combinations is again $n!$ and the value is
\be
-\frac{(-1)^n}{2}\sum_{n=1}^{\infty}{\int{\delta X(x)^n A(y)A(z)\Delta(x,y,z)}},
\ee
where
\be
\Delta(x,y,z)= \delta(x-y)+\frac{1}{2}\delta(y-z)+2\bigg( \int_{-\infty}^{+\infty}{\frac{dq}{q^2}} \bigg)\;\delta(x-y)\delta(x-z).
\ee
And the sum will be
\bea
&& -\frac{1}{2}\int{F(1+\delta X(x))A(x)A(y)dxdy} - \frac{1}{4}\int{F(1+\delta X(x))A(y)^2dxdy}\cr\cr
&&  -\bigg( \int_{-\infty}^{+\infty}{\frac{dq}{q^2}} \bigg) \int{F(1+\delta X(x))A(x)^2dx},
\eea
where
\be\label{deltafunction2}
F(x)=\frac{1}{x}-1,
\ee
which means we have
\bea\label{newren1}
&& -\frac{1}{2}\int{F(X(x))A(x)A(y)dxdy} - \frac{1}{4}\int{F(X(x))A(y)^2dxdy}\cr\cr
&& - \bigg( \int_{-\infty}^{+\infty}{\frac{dq}{q^2}} \bigg) \int{F(X(x))A(x)^2dx}.
\eea
From \ref{newren1} it is obvious that we need a new renormalization for $A^2$-term. For pattern 2 it is straightforward to show that the result is
\be
\frac{(-1)^{i+j}}{2}\int{\delta X(x)^iA(y)\delta X(z)^jA(w)\Delta(x,y,z,w)\;dxdydzdw}
\ee
with
\be
\Delta(x,y,z,w)=-\big[\delta(y-z)\delta(x-w)+\frac{1}{2}\delta(y-x)\delta(x-w)+\frac{1}{2}\delta(y-z)\delta(z-w)\big].
\ee
We can write the sum in this form
\be
W=\sum_{n=2}^{\infty}\sum_{i=1}^{n-1}(1+\delta_{ij})W(i,j),
\ee
where $j=n-i$ and
\bea
W(i,j)=\frac{(-1)^{i+j}}{4}\int{\delta X(x)^iA(y)\delta X(z)^jA(w)\Delta(x,y,z,w)\;dxdydzdw}
\eea
and the result is
\bea
W=&&\frac{1}{4}\int{F(1+\delta X(x))A(y)F(1+\delta X(z))A(w)\Delta(x,y,z,w)\;dxdydzdw}\cr\cr
&&+\frac{1}{2}\int{F(1-\delta X(x)\delta X(z))A(y)A(w)\Delta(x,y,z,w)\;dxdydzdw},
\eea
where $F$ is the function \ref{deltafunction2}. As before it means we should have
\bea
W=&&-\frac{1}{4}\int{ \big[ F(X(x))A(x) \big]\big[ F(X(y))A(y) \big]\;dxdy }\cr\cr
&&-\frac{1}{4}\int{ \big[ F(X(x))A(x)^2 \big]\big[ F(X(y)) \big]\;dxdy }\cr\cr
&&-\frac{1}{2}\int{ \big[ F\big(X(x)+X(y)-X(x)X(y)\big)A(x) A(y) \big]\;dxdy }\cr\cr
&&-\frac{1}{2}\int{ \big[ F\big(X(x)+X(y)-X(x)X(y)\big)A(x)^2 \big]\;dxdy }.
\eea
These terms do not contribute to the energy-momentum tensor, since they and their variations, at $X=1$, \emph{vanish}. For simplicity let us define these compact forms
\bea
&&F.A=\int{F(X(x))A(x)\;dx}\cr\cr
&&F.1=\int{F(X(x))\;dx}\cr\cr
&&A.1=\int{A(x)\;dx}.
\eea
And also
\bea
&&G(x,y)=F\big(X(x)+X(y)-X(x)X(y)\big)\cr\cr
&&f.G = \int{f(x)G(x,y)\;dx}\cr\cr
&&f.G.h=\int{f(x)G(x,y)h(y)\;dxdy}\cr\cr
&&G.1=\int{G(x,y)\;dy}.
\eea
So we can write the previous result in this form
\be
W=-\frac{1}{4}\big[(F.A)^2+(F.A^2)(F.1)+2\;(A.G.A)+2\;(A^2.G.1)\big].
\ee
And in general the $A^2$-term is
\bea
&&- \frac{1}{2} (F.A)(A.1) - \frac{1}{4}(F.1)(1.A^2) - \bigg( \int_{-\infty}^{+\infty}{\frac{dq}{q^2}} \bigg)(F.A^2)\cr\cr
&&-\frac{1}{4}\big[(F.A)^2+(F.A^2)(F.1)+2\;(A.G.A)+2\;(A^2.G.1)\big].
\eea
You can check that after exponentiating all the terms\footnote{Since they will be repeated in next terms, that we have not shown, as \emph{non-connected} diagrams and finally they will be exponentiated, as they should.}, \emph{the full renormalized effective action to second order in transverse momentums} is
\bea
&&\Gamma_{r-full}= \int{\frac{1}{\kappa^2}\big[\mathcal{R}-2\Lambda \big]d^{d+1}x}+\Gamma_r(g),
\eea
where
\bea\label{FullFinal}
\Gamma_r(g)=&&C_1\int{ln(X(x))\;dx}\cr\cr
+&&C_2\big[ (1.B)+(1.A^2)+(F.A^2) \big]\cr\cr
-&&\frac{1}{4}(F.1)\big[(1.B)+(1.A^2)\big]\cr\cr
-&&\frac{1}{2}(F.A)(1.A) -\frac{1}{4}(1.A)^2\cr\cr
-&&\frac{1}{4}\big[(F.A)^2+(F.A^2)(F.1)+2\;(A.G.A)+2\;(A^2.G.1)\big].
\eea
Remember that there is an overall integral over travnsverse momentums. Varing this effective action, when $X$ is not fixed, and then fixing the gauge $X=1$, produces all terms we have calculated so far. Here we just note that if we want to move the action out of the gauge point $X=1$, by a transformation from $\xi$ to another coordinate $\rho(\xi)$, the coupling $C_1$ transforms as a \emph{covariant vector}, since it is nothing but $\lim_{\xi \rightarrow \xi_0}\delta (\xi - \xi_0)$.

\section{Summary}
In this paper we have introduced and used a new method of calculating the gravitational effective action for planar geometries named \emph{gauge fixing method}.
we showed that the energy-momentum tensor to second order in transverse momentums is
\be
T_{AB}=H_{AB}-\frac{1}{2}\big( g^{MN}H_{MN} \big)g_{AB},
\ee
where $H_{AB}$ is
\bea
H_{\xi\xi}(x)&&=C_1-C_2A(x)^2+\frac{1}{2}A(x)\int{A(y)dy}+\frac{1}{4}\int{\big[A(y)^2+B(y)\big]dy}\cr\cr
H_{\xi\mu}(x)&&=i\;\big[C_2A(x) - \frac{1}{4}\int{A(y)dy}\big]k_\mu\cr\cr
H_{\mu\nu}(x)&&=C_2\;k_\mu k_\nu
\eea
and we calculated the effective action as a series in transverse momentums to second order
\bea
&&\Gamma(g)=\int{\Gamma(X,k)\;d^dk}\cr\cr
&&\Gamma(X,k) = \Gamma(X)+\Gamma^\mu (X)k_\mu + \Gamma^{\mu\nu} (X)k_\mu k_\nu + ... ,
\eea
where
\bea
\Gamma(X) &&= C_1\int{ln(X(x))\;dx}\cr\cr
\Gamma^\mu (X)&&=0 \cr\cr
\Gamma^{\mu\nu} (X)&&=\frac{1}{4}(1.A^\mu)(1.A^\nu)\cr\cr
+&&C_2\big[ (1.B^{\mu\nu})-(A^\mu .A^\nu)-(F.A^\mu .A^\nu) \big]\cr\cr
+&&\frac{1}{4}(F.1)\big[(A^\mu .A^\nu)-(1.B^{\mu\nu})\big]\cr\cr
+&&\frac{1}{4}\big[(F.A^\mu)(1.A^\nu) + (F.A^\nu)(1.A^\mu)\big]\cr\cr
+&&\frac{1}{4}\big[(F.A^\mu)(F.A^\nu) + (F.1)(F.A^\mu .A^\nu)\big]\cr\cr
+&&\frac{1}{2}\big[(A^\mu .G. A^\nu) + (A^\mu .A^\nu .G.1)\big].
\eea

\section{Discussion}
In this paper we have introduecd and used the \emph{gauge fixing method} to calculate the partition function and the vacuum expectation value of the energy-momentum tensor. This method can also be used for any other calculation of quantum field theory in curved planar space-times. The method we have presented here is just a new way of thinking about calculations in quantum field theory. Path integral formalism is a very clean way to express this viewpoint although the canonical formalism is also possible. There are some points:

\emph{Boundary conditions} - In the calculation we have done, there is not any boundary condition. Imposing boundary conditions in $\xi$ direction changes the spectrum and the Hilbert space of the theory. The Green's function will be changed and finally the results of the calculation will be changed as well. The boundary conditions depend on the problem in hand but, the general method is again the same as we used here.

\emph{Non-perturbative calculation} - In this paper we used a perturbative method to calculate the effective action and energy-momentum tensor but, we can also use non-perturbative methods. For example consider the full, $d+1$-dimensional, Green's function $\GG_f$. Originally for calculating the Green's function we need to have the \emph{full spectrum} of the \emph{full operator} $\p_\xi ^2-A(\overleftarrow{\p}-\overrightarrow{\p})+B$. Then the Green's function is
\be
\GG_f(x,y) = -\sum_n{\frac{\phi_n^*(x)\phi_n(y)}{\lambda_n^2}},
\ee
where $-\lambda_n^2$ are the eigenvalues of the operator. But we do not have the full spectrum of the full operator. Specially when the metric is not a specific one, like the case of gravitational effective action, this calculation is impossible \emph{in principle}! On the other hand in gauge fixing method It is so easy to check that the Green's function satisfies the equation
\bea
&&\GG_f(x,y)=\GG(x,y)-\int{\GG(x,z)B(z)\GG_f(z,y)\;dz}\cr\cr
&&-\int{\big[\GG(x,z) A(z) \p_z\GG_f(z,y) - \p_z\GG(x,z) A(z) \GG_f(z,y)\big]\;dz}.
\eea
This is an integral equation and if we can solve it, then we will have the full Green's function \emph{non-perturbatively}.

\emph{Observer} - As mentioned previousely our method has two steps of gauge fixing. In first step we restrict ourselves to observers that in their frames the metric is a function of just one coordinate. In second step we change the coordinate $z$ to $\xi$. This coordinate transformation is just a change in the way that the observer measures the lengths along $\xi$ direction and the observer is the same as before. In summary in first step we restrict the general observer to a special sort of observers and in second step there is no more restriction on the observers that we have fixed in previous step.

\emph{Interactions} - Including interaction terms like $\lambda\phi ^4$, is also possible in this formalism but leads to mixing between transverse momentums and so there is not an overall integral over transverse momentums in the final result. But interesting phenomena arise from quantizing matter field on curved space and interaction terms just modify the numbers, but not the phenomena themselves.

\emph{More than one coordinate} - Present method is applicable to cases that the metric is a function of just one coordinate. It will be very interesting and useful if we can develop a method for cases that the metric depends on more than one coordinate. It of course needs more investigations.

\vspace{.5cm}
\emph{Acknowledgement} - I would like to thank A. Davody and M. Safari for useful discussions.


\end{document}